\documentclass[preprint2]{aastex} 
\usepackage{graphicx}
\usepackage{amsmath}
\usepackage[colorlinks=true,citecolor=blue,breaklinks=true,linktocpage=true]{
hyperref}
\bibpunct{(}{)}{;}{a}{}{,} 
\usepackage[switch,pagewise]{lineno}
\usepackage{xspace}
\usepackage{xparse} 

\newcommand{\alfven}{Alfv\'en\xspace}
\newcommand{\unit}[1]{\ensuremath{\,\mathrm {#1}}}  
\newcommand{\A}{\ensuremath{\text{A}}}%
\newcommand{\s}{\ensuremath{\text{s}}}

\renewcommand{\i}{\ensuremath{\text{i}}}  
\newcommand{\e}{\ensuremath{\text{e}}} 
\DeclareDocumentCommand\figref{ m g }{{Figure~\ref{#1}\IfNoValueF {#2} {(#2)}}}
\newcommand{\secref}[1]{Section~\ref{#1}}

\renewcommand{\eqref}[1]{Equation~\ref{#1}}
\newcommand{\tabref}[1]{Table~\ref{#1}}
\renewcommand{\deg}{\ensuremath{^\circ}}
\newcommand{\RN}[1]{\uppercase\expandafter{\romannumeral#1}}
\newcommand{\mtilde}{\ensuremath{\raise.17ex\hbox{$\scriptstyle\sim$}}}
\newcommand{\kmps}{\ensuremath{\unit{km\,s^{-1}}}}

\newcommand{\ev}{Event~V\xspace}
\newcommand{\mv}{Model~V\xspace}
\newcommand{\arv}{AR 11283\xspace}
\newcommand{\datev}{06 Sep 2011\xspace}

\newcommand{\ea}{Event~A\xspace}
\newcommand{\ma}{Model~A\xspace}
\newcommand{\ara}{AR 11112\xspace}
\newcommand{\datea}{16 Oct 2010\xspace}

\newcommand{\ew}{Event~W\xspace}
\newcommand{\mw}{Model~W\xspace}
\newcommand{\arw}{AR 11121\xspace}
\newcommand{\datew}{03 Nov 2010\xspace}
\defcitealias{yuan2015sv}{Paper I} 	
\newcommand{\sdo}{{\it SDO}\xspace}
\newcommand{\SDO}{{\it Solar Dynamics Observatory}}

\shorttitle{standing kink wave}
\shortauthors{Yuan et al.}

\begin{document}

\title{Forward Modelling of Standing Kink Modes in Coronal Loops \RN{2}.
Applications}
\author{Ding Yuan\altaffilmark{1,2,3}}
\email{DYuan2@uclan.ac.uk}
\author{Tom {Van Doorsselaere}\altaffilmark{1}}

\altaffiltext{1}{Centre for mathematical Plasma Astrophysics, Department of
Mathematics, KU Leuven, Celestijnenlaan 200B bus 2400, B-3001 Leuven, Belgium}
\altaffiltext{2}{Jeremiah Horrocks Institute, University of Central Lancashire,
Preston PR1 2HE, UK}
\altaffiltext{3}{Key Laboratory of Solar Activity, National Astronomical
Observatories, Chinese Academy of Sciences, Beijing, 100012, China}
\begin{abstract}
Magnetohydrodynamic waves are believed to play a significant role in coronal
heating, and could be used for remote diagnostics of solar plasma. Both the
heating and diagnostic applications rely on a correct inversion (or backward
modelling) of the observables into the thermal and magnetic structures of the
plasma. However, owing to the limited availability of observables, this is an
ill-posed issue. Forward Modelling is to establish a plausible mapping of plasma
structuring into observables. In this study, we set up forward models of
standing kink modes in coronal loops and simulate optically thin emissions in
the extreme ultraviolet bandpasses, and then adjust plasma parameters and
viewing angles to match three events of transverse loop oscillations observed by
the \SDO/Atmospheric Imaging Assembly. We demonstrate that forward models could
be effectively used to identify the oscillation overtone and polarization, to
reproduce the general profile of oscillation amplitude and phase, and to predict
multiple harmonic periodicities in the associated emission intensity and loop
width variation.
\end{abstract}

\keywords{Sun: atmosphere --- Sun: corona --- Sun: oscillations ---
magnetohydrodynamics (MHD) --- waves}

\section{Introduction}
\label{sec:intro}
The solar atmosphere and its magnetic structure support a variety of
magnetohydrodynamic (MHD) wave phenomena \citep[see reviews by
][]{nakariakov2005,demoortel2012,jess2015}.
Standing kink waves in coronal loops \citep{edwin1983,goossens2014} were first
observed by the Transition Region and Coronal Explorer
\citep{nakariakov1999,aschwanden1999}. Coronal loops are observed to oscillate
transversely in response to explosive events, i.e., mass ejections
\citep{schrijver2002, zimovets2015}, filament destablizations
\citep{schrijver2002}, magnetic reconnections \citep{he2009}, or vortex shedding
\citep{nakariakov2009}. This kind of transverse loop oscillations has typical
amplitude of the order of the loop radius and period at minute timescales, and
is damped within several wave cycles \citep{aschwanden2002}. Another type of
low-amplitude (sub-megameter scale) transverse oscillations is observed to last
for dozens of wave cycles without significant damping
\citep{nistico2013,anfinogentov2013,anfinogentov2015}; no apparent exciter has
been identified for this type of coronal oscillations.

The discovery of standing kink mode initiated a new field, MHD seismology
\citep[remote diagnostics of solar plasma,][]{nakariakov2005,demoortel2012}.
\citet{nakariakov2001} inferred the magnetic field strength of coronal loops
using the wave parameters. Subsequent applications spread to studying the
cross-sectional loop structuring \citep{aschwanden2003}, \alfven transit times
\citep{arregui2007}, polytropic index and heat transport coefficient
\citep{vandoorsselaere2011}, magnetic topology of sunspots
\citep{yuan2014lb,yuan2014cf}, magnetic structure of large-scale streamers
\citep{chen2010,chen2011}, and the correlation length of randomly structured
plasmas \citep{yuan2015rs}.

\citet{demoortel2009} made the first attempt to validate MHD seismology with a
three-dimensional (3D) MHD simulation, and found that the magnetic field
strength obtained by MHD seismology is only half of the input value.
\citet{pascoe2014} demonstrated that, if a loop is excited by an external
driver, a second period would blend with the eigenmode and may mislead the
estimation of wave period. \citet{aschwanden2011} and \citet{verwichte2013}
demonstrated that the magnetic field inferred by MHD seismology only agrees with
the potential field model within a factor of about two. \citet{chen2015} found
that the magnetic field inverted with a kink MHD mode agrees with the input
average field along a coronal loop. Therefore, forward modelling is required to
establish the connectivity between the plasma structuring and the spectrographic
and imaging observables \citep[e.g.,][]{yuan2015fm,antolin2013}.
\citet{wang2008} applied a simple geometric model to identify the polarizations
and the longitudinal overtones of kink waves observed at various parts of the
solar disk. \citet{yuan2015sv} (referred to as \citetalias{yuan2015sv}
hereafter) synthesised the spectrographic observations of the standing kink
modes of coronal loops and demonstrated that the quadrupole terms in the kink
mode solution could lead to the detection of rotational motions and nonthermal
broadening at loop edges, and emission intensity and loop width variation.

In this paper, we apply the forward modelling of \citetalias{yuan2015sv} to
interpret observational data. \secref{sec:models} presents the selection of
\SDO/Atmospheric Imaging Assembly \citep[\sdo/AIA][]{lemen2012,boerner2012}
observations and the corresponding forward models. \secref{sec:app} demonstrates
how forward modelling could be applied to quantify the kink wave amplitude,
explain the loop width oscillation and identify the overtones. finally, the
conclusion is given in \secref{sec:con}.

\section{Observations and forward models}
\label{sec:models}
In this study, we select three events of kink loop oscillations
(\tabref{tab:events}) and constructed the relevant forward models
(\tabref{tab:models}) based on measured parameters. The selected events were
previously analyzed in \citet{verwichte2013}, \citet{aschwanden2011} and
\citet{white2012b}, respectively. Henceforth, we refer to them as Event V, A,
and W, respectively; while the associated models are labeled as Model V, A, and W. We only constructed the fundamental  mode for Event V and A; whereas for Event W, \textbf{we simulate the $1^\mathrm{st}$, $2^\mathrm{nd}$ and $3^\mathrm{rd}$ overtones, i.e., W$_1$, W$_2$
and W$_3$ (\tabref{tab:models}). We could exclude the possibility of the fundamental mode, as already did in \citet{white2012b}, so we only include the illustrations and discussion for $n=2$ and $3$ overtones.} Here the $n$-th overtone means that there is $n$ nodes in a standing
wave. $n=1, 2, 3$ stands for the fundamental, 2nd and 3rd overtones,
respectively. 

For each event, we configure a straight, magnetized plasma cylinder and
its ambient plasma using observed parameters. Then we solve the analytical model for the kink
MHD mode \citep[see e.g., ][]{edwin1983,goossens2014}, the wave amplitude is
choosen as estimated in observations. Three dimensional distributions of plasma
density, temperature and velocity are passed to a Forward Modelling code
\citep[FoMo\footnote{The FoMo code is available at
\url{https://github.com/TomVeeDee/FoMo}}, Van Doorsselaere et al., Frontiers, submitted;][]{yuan2015fm}. The FoMo
code includes the atomic emission effect in the optically thin plasma approximation and
synthesizes spectrographic and imaging observations. Details on modelling
standing kink wave are given in \citetalias{yuan2015sv}. In this study, the AIA imaging observation of standing kink waves are synthesized to match observations by varying the viewing angles.

In this paper, we present the methods to identify polarisation and overtones of standing kink modes (\secref{sec:mode}), the properties inferred from the
amplitude and phase distribution (\secref{sec:amp}), and the periodicity in loop
intensity and width variations (\secref{sec:width}).

\subsection{Event and Model V}
\begin{figure}[ht]
\centering
\includegraphics[width=0.5\textwidth]{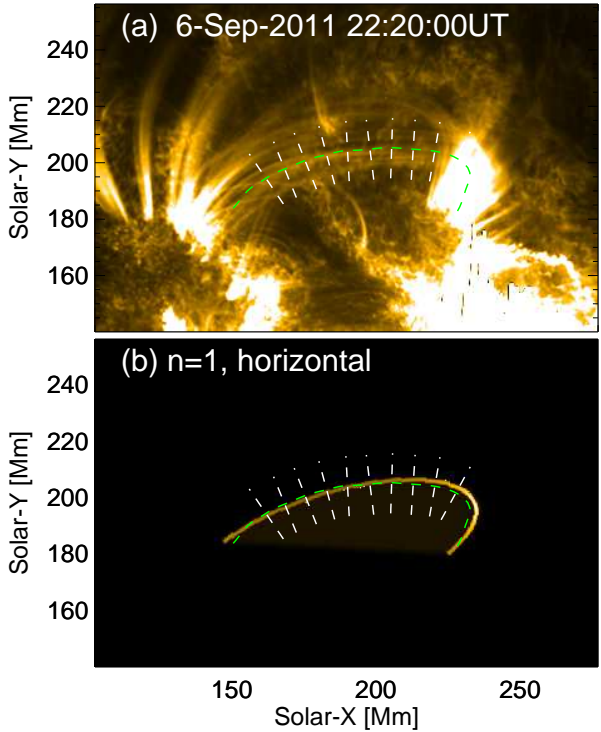}
\caption{(a) FOV of \arv (\ev) observed in the AIA 171 \AA{} channel on \datev.
(b) Synthetic view (\mv) in the 171 \AA{} bandpass. The green dashed lines label
the loop of interest (loop coordinate increases counter-clockwisely); while the
set of white dashed segments denotes the slits used for time distance plots
(\figref{fig:xt_v}). \label{fig:fov_ev}}
\end{figure}

\ev \citep{verwichte2013} was observed at \arv in the AIA 171 \AA{} channel on
\datev .
\arv was associated with a Hale-class $\beta\gamma$ or $\beta\delta$ sunspot;
the general $\beta$ (bipolar) magnetic configuration formed a bundle of distinct
coronal loops connecting the opposite polarities. It crossed the central
meridian on the previous day and was well exposed for AIA observation on \datev
(\figref{fig:fov_ev}). Two or more loops oscillated transversely in response to
a GOES class X2.1 flare, which started at 22:12 UT and peaked at 22:20 UT. A
fainter loop (labeled by the green dashed line in \figref{fig:fov_ev},
corresponding to loop \# 2 in \citet{verwichte2013}) oscillated for about four
wave cycles. It did not fade out, nor become significantly brighter during the
kink oscillation, and therefore, it is chosen for further investigation. In our study, the latter three wave cycles were selected for modelling.

\citet{verwichte2013} performed 3D stereoscopy and gave a loop geometry with a length of $L_0=160\unit{Mm}$ and a radius of $a=0.85\unit{Mm}$. The plasma temperature was assumed to be the nominal response temperature ($0.8\unit{MK}$) of the AIA 171 \AA{} bandpass, since this loop was only visible in this channel \citep{verwichte2013}. The electron density was estimated at a lower limit at $n_{e\i}=0.7\cdot10^9\unit{cm^{-3}}$. The loop oscillated with a period of about $2.0\unit{min}$ and an amplitude at $1.9\pm1.0\unit{Mm}$ (about $1.0a$-$3.4a$). The relevant measurements are summarised in \tabref{tab:events} (\ev).

We model this loop with a length of $L_0=160\unit{Mm}$ and a radius of
$a=0.85\unit{Mm}$. The internal electron density is set to
$n_{e\i}=1.0\cdot10^{9}\unit{cm^{-3}}$, while the density ratio is defined as
$n_{e\i}/n_{e\e}=3.0$. The loop is filled with plasma at $T_\i=0.8\unit{MK}$,
1.5 times hotter than the ambient plasma. We used $B_\i=B_\e=30\unit{G}$ for
both internal and external magnetic field strength\footnote{$B_\e$ is
$0.07\unit{G}$ stronger than $B_\i$ according to the calculation using total
pressure balance, however, we round the numbers to two significant digits in
this paper.}. This equilibrium state has internal acoustic and \alfven speeds
$C_{\s\i}=150\unit{km\,s^{-1}}$ and $V_{\A\i}=2100\unit{km\,s^{-1}}$, and
$C_{\s\e}=120\unit{km\,s^{-1}}$ and $V_{\A\e}=3600\unit{km\,s^{-1}}$ as the
corresponding external speeds. The oscillation period is about $126\unit{s}$;
and the amplitude $\xi_0$ about $2.0\unit{Mm}$ ($2.4a$).

The horizontal mode is modelled with parameters listed in \tabref{tab:models}
(\mv). The viewing angle $[30\deg,130\deg]$ \citepalias[see][for
definition]{yuan2015sv} matches the loop orientation very well
(\figref{fig:fov_ev}). The synthetic image is interpolated into AIA resolution
and aligned by matching the centre of the baseline at
$[226.3\arcsec,215.3\arcsec]$. Then the aligned synthetic image is then rotated by an angle of $3\deg$ clockwisely. 

\begin{figure*}[ht]
\centering
\includegraphics[width=0.40\textwidth]{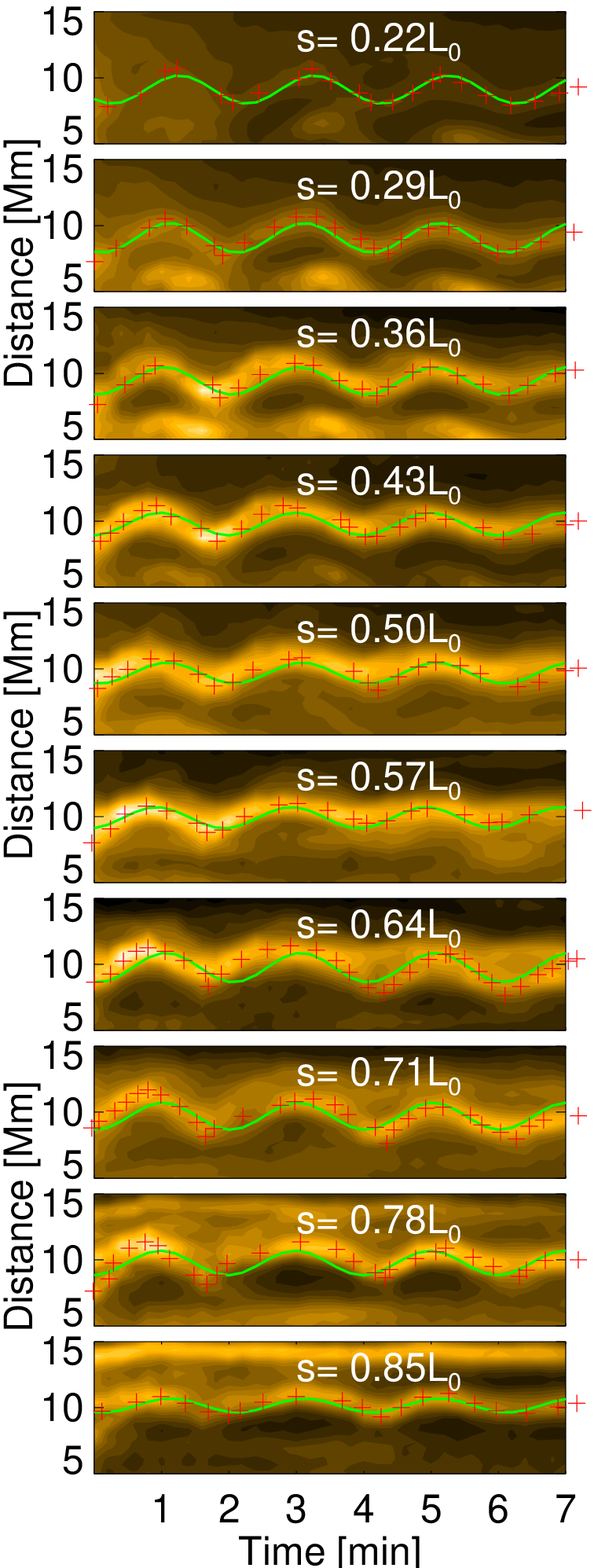}
\includegraphics[width=0.40\textwidth]{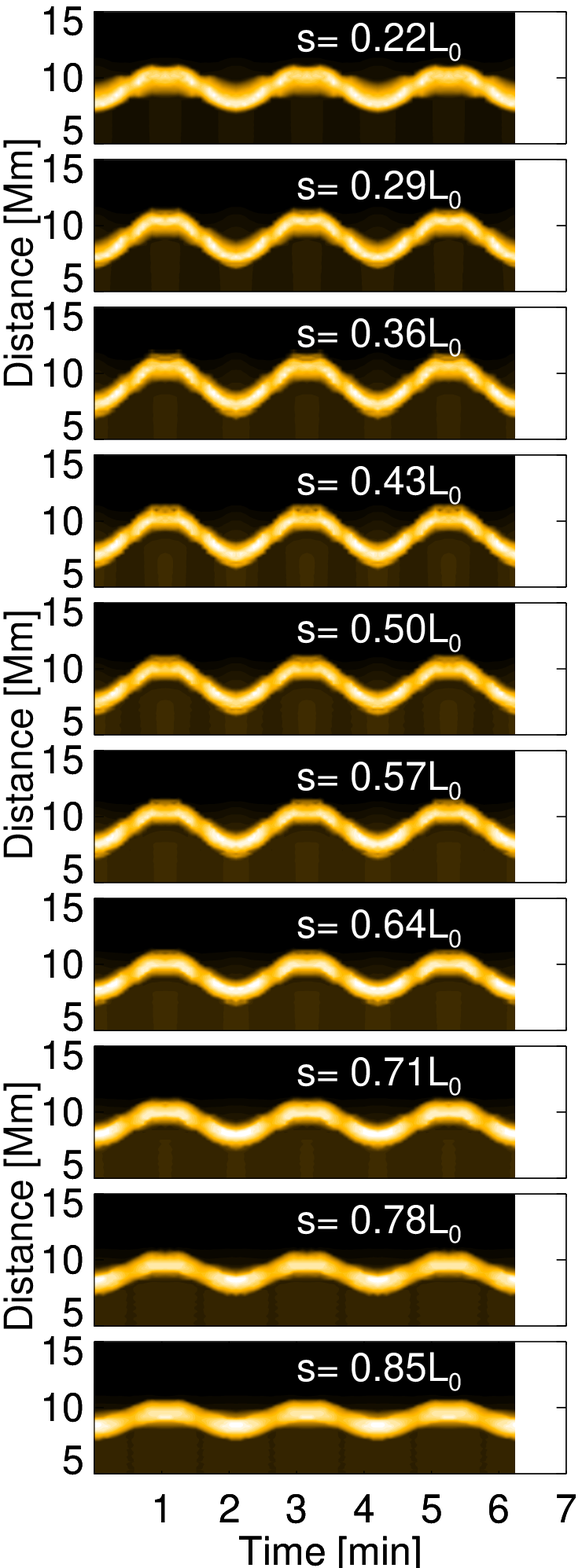}
\caption{Left panel: time distance plots along slits normal to loop spine at
selected loop coordinates in \ev. The crosses are the traced loop motions, while
the green continuous line are the sinusoidal fits. Right panel: same as left
panel, but extracted at the synthetic view for the $n=1$ horizontal mode
(\mv).\label{fig:xt_v}} 
\end{figure*}

\begin{figure}[ht]
\centering
\includegraphics[width=0.5\textwidth]{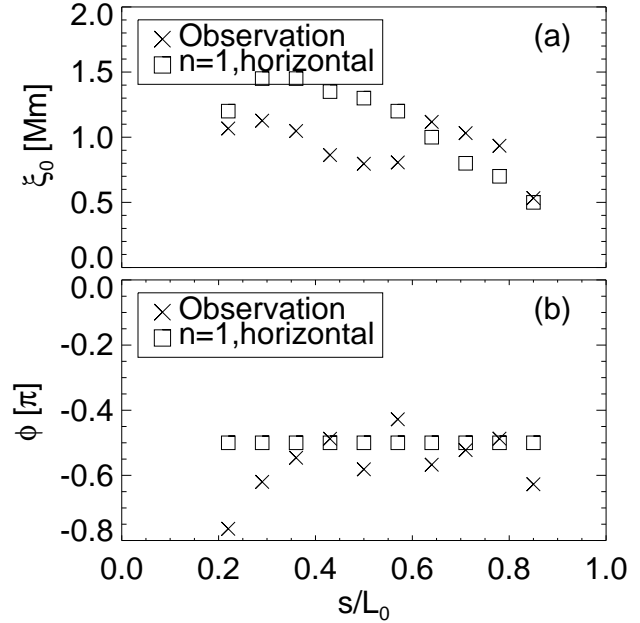}
\caption{Oscillation amplitude (a) and phase (b) as functions of the loop
coordinate, measured in \ev and the $n=1$ horizontal mode (\mv),
respectively.\label{fig:ph_ev}}
\end{figure}

\figref{fig:xt_v} displays the time distance plots along the slits labeled in
\figref{fig:fov_ev} (in counter-clockwise order). The oscillations at various
parts of the loop are in phase with each other and exhibit amplitude variation
along the loop. The loop motions are traced manually (red crosses in
\figref{fig:ph_ev}), and then, the time series of loop displacement was fitted
with a sinusoidal function, as presented in \citet{aschwanden2011} but without
the damping term. \figref{fig:ph_ev} plots the fitted amplitude and phase for
\ev. The same procedure is applied to both \ea and \ew. \textbf{We also measure the amplitude and phase from the synthetic time-distance plots as diplayed in \figref{fig:xt_v}, \ref{fig:xt_a}, and \ref{fig:xt_w}, and plots them in \figref{fig:ph_ev}, \ref{fig:ph_ea}, and \ref{fig:ph_ew}, respectively. In the synthetic time-distance plots, we simply track loop positions by finding the maximum intensity within each time step.}

The selected loop in \ev is clear from background contamination,
therefore, we measure the oscillation along the slit at $s=0.5L_0$ in detail. 
We fit Gaussian functions to the intensity profiles across the loop at $s=0.5L_0$ and extract the loop displacement, flux, and width (full width at half maximum, FWHM) variations. And then we compare them with synthetic kink oscillation, see results in \secref{sec:width}.

\subsection{Event and Model A}
\begin{figure*}[ht]
\centering
\includegraphics[width=0.8\textwidth]{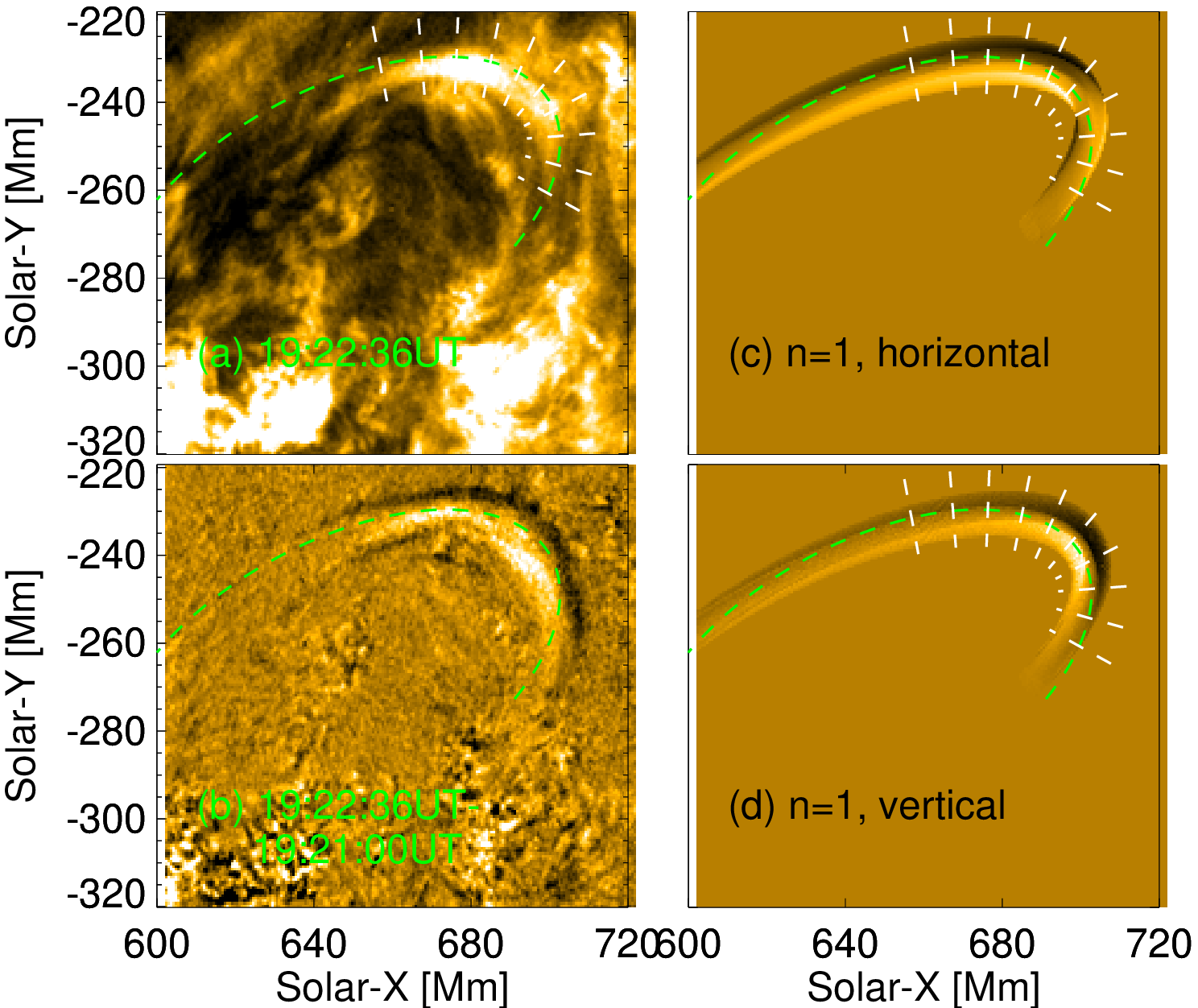}
\caption{(a) FOV of \ea observed in the \sdo/AIA 171 \AA{} channel. (b)
Difference image displaying the loop oscillation. (c) Difference image of the
$n=1$ horizontal kink mode in the 171 \AA{} bandpass. (d) Same as (c), but for
the $n=1$ vertical mode. The green dashed lines in all panels label the
approximate loop spine, the loop coordinate increases counter-clockwisely; while
time-distance plots (\figref{fig:xt_a}) are extracted along the set of white
dashed slits. \label{fig:fov_ma}}
\end{figure*}

\begin{figure*}[ht]
\centering
\includegraphics[width=0.32\textwidth]{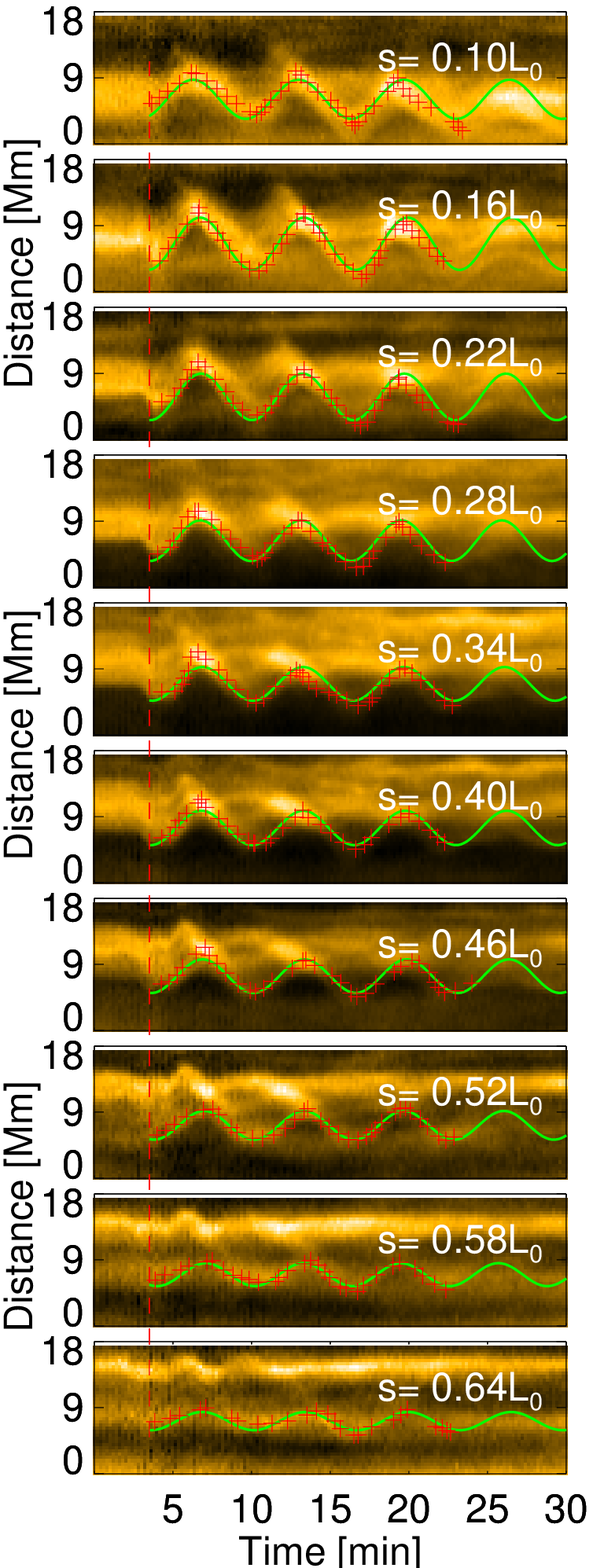}
\includegraphics[width=0.32\textwidth]{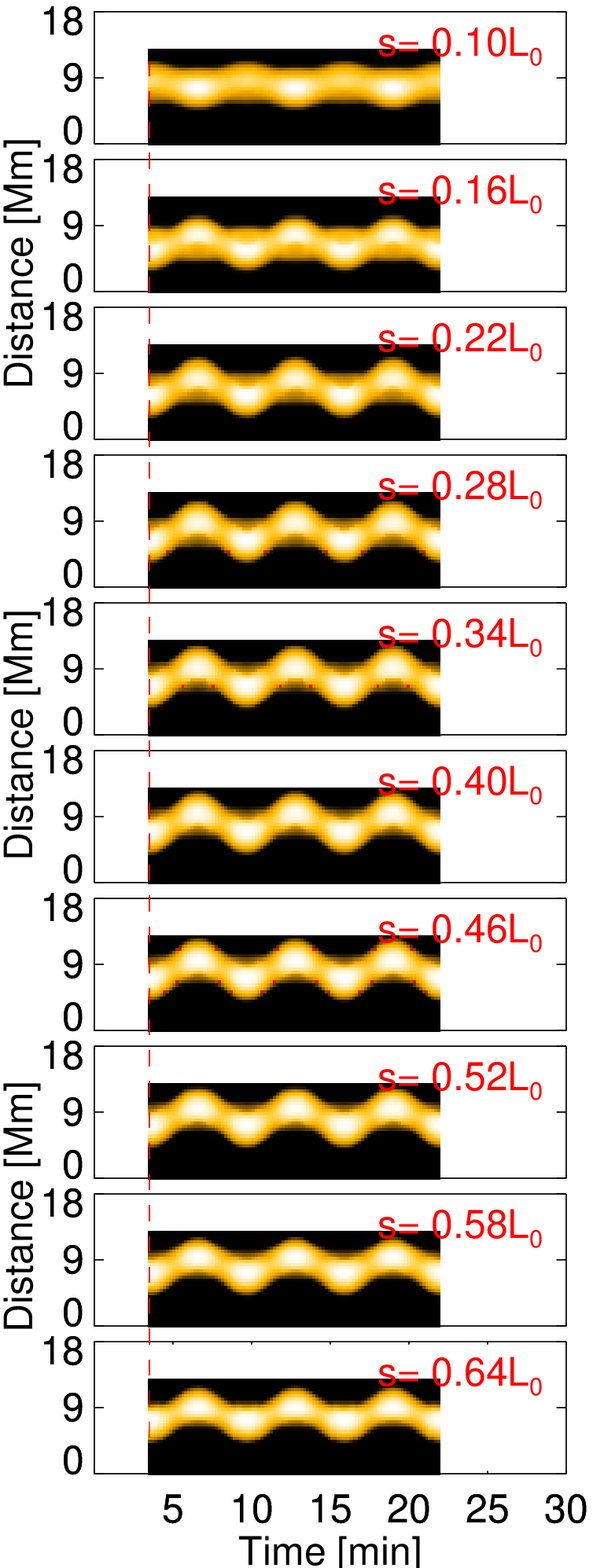}
\includegraphics[width=0.32\textwidth]{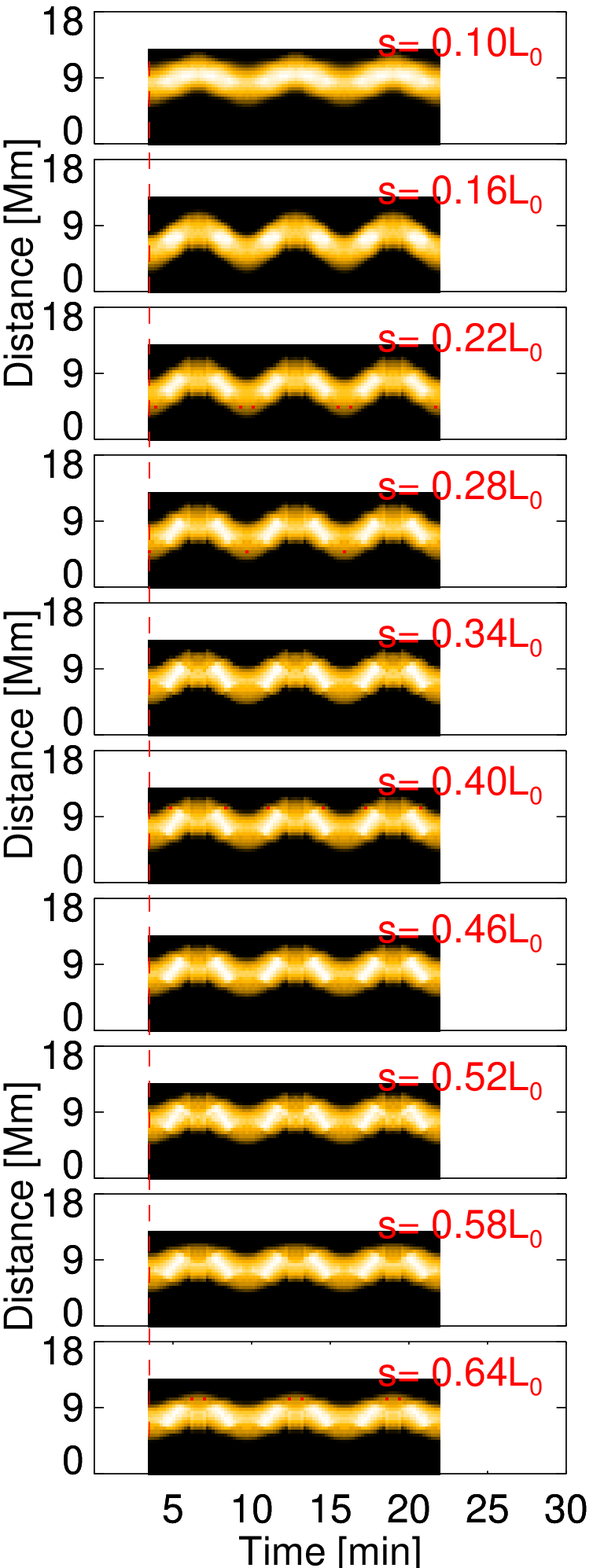}
\caption{Time distance plots at selected loop coordinates extracted along the
slits in the \sdo/AIA 171 \AA{} images during \ea (left column) and synthetic
emission images for the $n=1$ horizontal (middle column) and vertical modes
(right column). The red dashed line marks the start of oscillations; the crosses label the identified loop displacement; and the yellow continuous lines are the
sinusoidal fits. The time starts at 19:10:00 UT \datea. \label{fig:xt_a}}
\end{figure*}

\ea was studied in detail by \citet{aschwanden2011}. On \datea, a GOES-class
M2.9 flare occurred at active region \ara. The excited coronal wave was observed
to propagate to the north-west of \ara and swept across the extended flare
ribbons \citep{kumar2013b}. A bundle of coronal loops was located at a distance
of about $0.32R_\sun$ to the disk centre (about $230\unit{Mm}$ north-west to
\ara, \figref{fig:fov_ma}). Sequential brightening of the flare ribbons provided a good estimate of the \alfven transit time, and therefore, the external \alfven speed of the loop system was roughly measured \citep{aschwanden2011}. Two or more adjacent loops oscillated for about three to four cycles, no significant
damping was observed. Moreover, the loop displacement appeared to exhibit a
saw-tooth pattern (\figref{fig:xt_a}), rather than sinusoidal curves as usually observed \citep{aschwanden2002}.

\ea was claimed to be a vertical transverse loop oscillation observed by the AIA 171 \AA{} channel \citep{aschwanden2011}. The loop length measured
$163\unit{Mm}$; and the radius about $2.5\pm0.3\unit{Mm}$. A bundle of loops
connected two opposite polarities that were not associated with any sunspots. A potential field extrapolation gave a field strength of about $6\unit{G}$ at the loop apex; while the field value was measured at $4.0\pm0.7\unit{G}$ using MHD seismology \citep{aschwanden2011}. The loop was filled with plasma of density of about $2\cdot10^{8}\unit{cm^{-3}}$ and temperature of about $0.6\unit{MK}$. The oscillation period was about $6.3\unit{min}$; and the amplitude about $1.7\pm0.4\unit{Mm}$. The measured parameters are listed in \tabref{tab:events} (\ea). 
  
We model the loop with a semi-toroidal geometry of a length at
$L_0=160\unit{Mm}$ and a radius at $a=2.5\unit{Mm}$. The internal and external
magnetic field is $B_\i=4.0\unit{G}$ and $B_\i=4.1\unit{G}$, respectively. The
loop is filled with plasma of density at $2.2\cdot10^{8}\unit{cm^{-3}}$, 12
times denser than the ambient plasma. The loop temperature is set at
$0.57\unit{MK}$, 1.5 times hotter than the background. The plasma $\beta$ is
about 0.054 and 0.003 for the internal and external plasma, respectively. This
configuration gives a kink mode solution with  $P_0=6.7\unit{min}$, obtained by
solving the dispersion relationship \citepalias[Equation 13 in][]{yuan2015sv}.
The oscillation amplitude is set at $\xi_0=4.5\unit{Mm}$ 
($1.8a$). 

We construct both horizontal and vertical kink wave models for this loop
(\tabref{tab:models}, \ma). The best matching viewing angle is
$[32\deg,135\deg]$. The centre of the baseline is placed at
$[646.2\arcsec,-274.8\arcsec]$; the synthetic image is interpolated in to AIA
resolution and rotated by an angle of $5\deg$ clockwisely.

\begin{figure}[ht]
\centering
\includegraphics[width=0.5\textwidth]{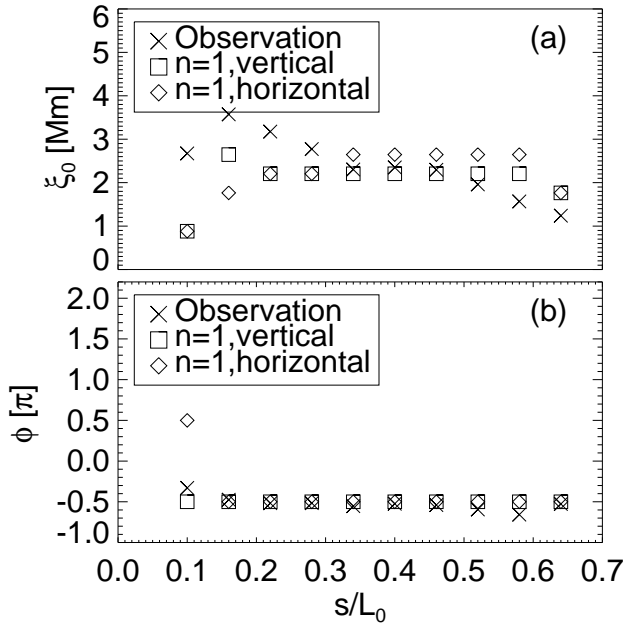}
\caption{Oscillation amplitude (a) and phase (b) as functions of the loop
coordinate, measured in \ea, the synthesised horizontal and vertical modes
(\ma), respectively. \label{fig:ph_ea}}
\end{figure}

\figref{fig:xt_a} illustrates the time distance plots along selected slits
normal to the loop spine (\figref{fig:fov_ma}). The loop oscillations are
coherently in phase along the loop, which confirms that the kink oscillation is
an established eigenmode of the loop. The synthetic kink wave exhibits similar
motions (\figref{fig:xt_a}, middle and right columns). The horizontal mode finds
intensity maxima when the loop oscillates to extreme positions, while the
vertical mode reaches maxima when the loop crosses the equilibrium position. The
phase and amplitude of the oscillation as functions of loop coordinates are
measured and plotted in \figref{fig:ph_ea}.

\subsection{Event and Model W}

\begin{figure*}[ht]
\centering
\includegraphics[width=\textwidth]{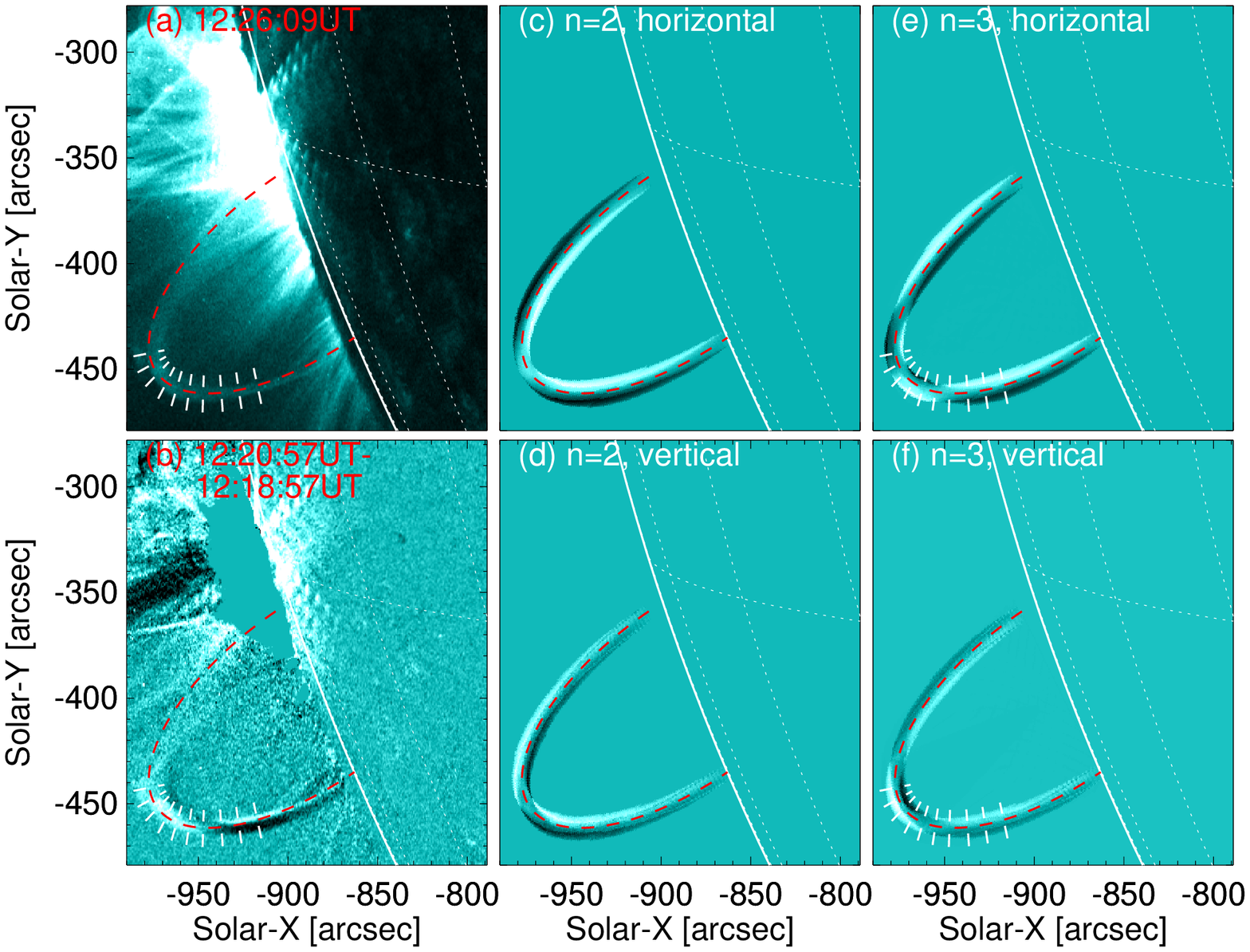}
\caption{(a) FOV of \arw observed at the South-East solar limb by the \sdo/AIA
131 \AA{} channel. The dashed line labels the hot flaring loop of interest. (b)
Difference images made by subtracting two images taken at about half an
oscillation cycle apart. (c) - (f) Difference images of two synthetic images
taken at half a wave cycle apart for $n=2$ horizontal and vertical overtones,
and $n=3$ horizontal and vertical overtones, respectively.
\label{fig:fov_rw}}
\end{figure*}

\arw was associated an Hale-class $\alpha$ sunspot group with a unipolar
magnetic configuration, observed on the east limb of the solar disk on \datew. A
GOES-class C3.4 flare started at about 12:12 UT and excited two EUV waves
\citep[see, e.g,][]{liu2014} or wave trains \citep[e.g.,][]{yuan2013fw}. A
magnetic flux tube that connected the main spot and another polarity was quickly
filled up with hot and dense plasma (\figref{fig:fov_rw}). A kink loop
oscillation was excited by the mass ejecta and exhibited non-coherent motions.
\ew was a sporadic transverse oscillation of a flaring loop observed in the
\sdo/AIA channels that are sensitive to hot plasma emissions (131 \AA{},
$\mtilde10\unit{MK}$). The loop supported possible higher longitudinal
overtones, and perhaps vertical polarisation of a kink wave \citep{white2012b}. 

\citet{white2012b} performed 3D stereoscopy loop reconstruction combining the
STEREO-B EUVI 195 \AA{} bandpass and the \sdo/AIA 131 \AA{} channel. This
procedure, using a low ($\mtilde1.6\unit{MK}$) and a high ($\mtilde10\unit{MK}$)
temperature channel, may have overestimated the loop length by a factor of two,
therefore, we measure the loop length by fitting a projected semi-torus to the
loop (\figref{fig:fov_rw}{a}), and obtain a loop length of $L_0=240\unit{Mm}$.
Differential emission measure (DEM) analysis using the forward-fitting technique
\citep{aschwanden2013} gives the loop radius $a=3.8\unit{Mm}$, electron density
$n_{e\i}=3.2\cdot10^9\unit{cm^{-3}}$, and loop temperature $T_\i=10\unit{MK}$.
The loop oscillated back and forth about every 5 min with an amplitude of about
$4.7\unit{Mm}$ ($1.2a$).

To identify the overtone number, we constructed models of $n=2$ and $3$
overtones with options of either vertical and horizontal polarization
(\figref{fig:fov_rw}). For the $n=2$ overtone, we use $B_\i=11\unit{G}$ and
$B_\e=17\unit{G}$ as internal and external magnetic field strength. The flux
tube is filled with plasma of $n_{e\i}=2.5\cdot10^9\unit{cm^{-3}}$ and
$T_\i=10\unit{MK}$, 4 times denser and 2 times hotter than the background.
Therefore, the internal and external plasma $\beta$ are about $1.7$ and $0.08$,
respectively, which are reasonable values for flaring loops \citep[see,
e.g.,][]{vandoorsselaere2011b}. In this configuration, the internal acoustic and
\alfven speeds are $C_{\s\i}=520\kmps$ and $V_{\A\i}=480\kmps$, while the
external speeds are $C_{\s\e}=370\kmps$ and $V_{\A\e}=1600\kmps$, respectively.
The kink mode solution gives the period $P_0=301\unit{s}$; the amplitude is set
at $\xi_0=1.5\unit{Mm}$ ($0.5a$). 

For the $n=3$ overtone, the internal and external magnetic field values are
$B_\i=9\unit{G}$ and $B_\e=15\unit{G}$, respectively. The loop is filled with
plasma at $n_{e\i}=3.0\cdot10^9\unit{cm^{-3}}$ and $T_\i=10\unit{MK}$, 2 times
denser and 2 times hotter than the ambient plasma. Then, the typical speeds are
$C_{\s\i}=520\kmps$, $V_{\A\i}=360\kmps$, $C_{\s\e}=370\kmps$, and
$V_{\A\e}=870\kmps$; and the internal and external plasma beta are 2.5 and 0.22,
respectively. $P_0=277\unit{s}$ is the period of the kink mode solution. We used
an oscillation amplitude at $\xi_0=1.5\unit{Mm}$ ($0.5a$). 

The synthetic images of both $n=2$ and $n=3$ modes are interpolated into AIA
resolution and aligned with AIA FOV by matching the centre of the loop baseline
at $[-884.7\arcsec,-392.9\arcsec]$; then they are rotated by an angle of
$-120\deg$ clockwisely.

\begin{figure*}[ht]
\centering
\includegraphics[width=0.32\textwidth]{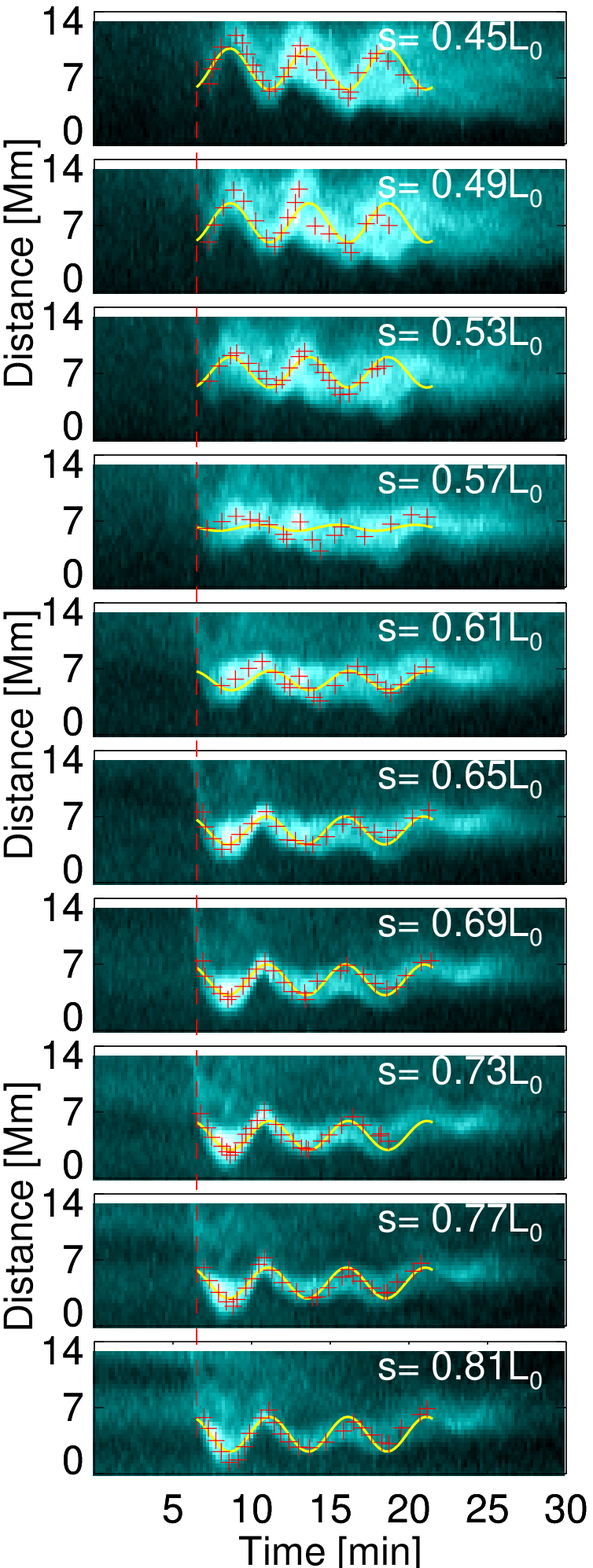}
\includegraphics[width=0.32\textwidth]{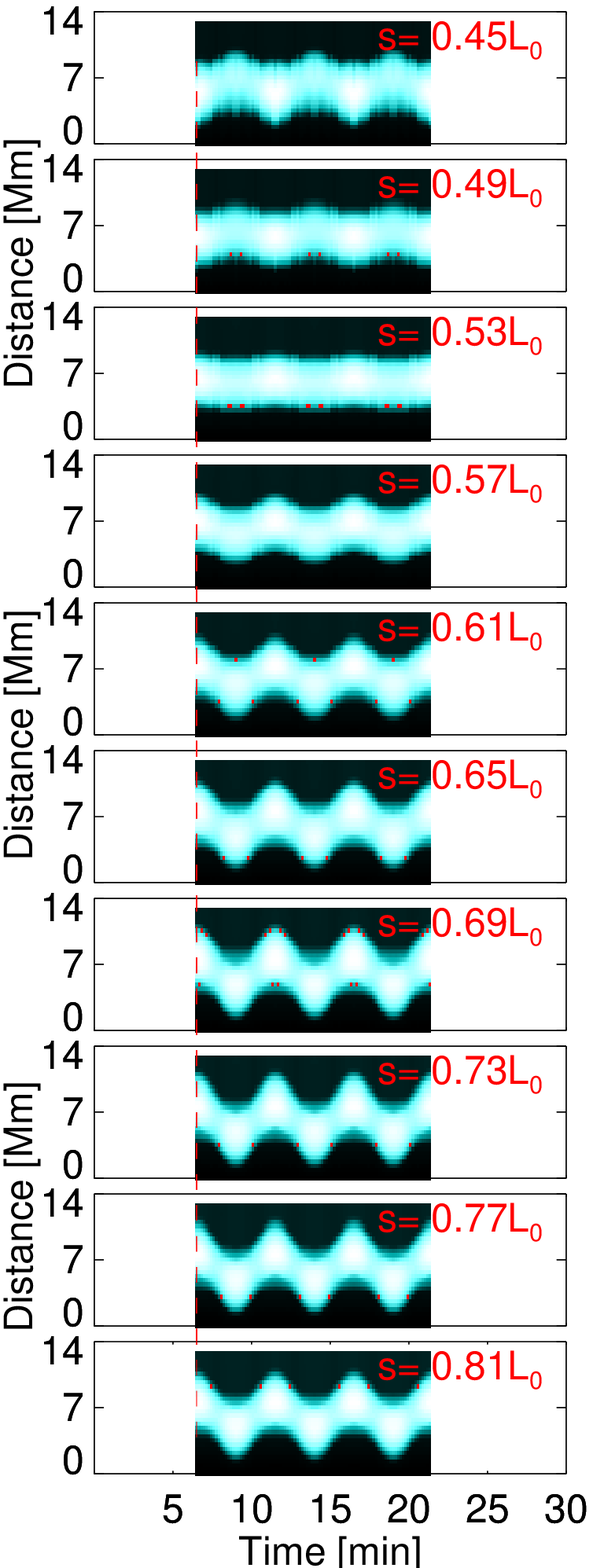}
\includegraphics[width=0.32\textwidth]{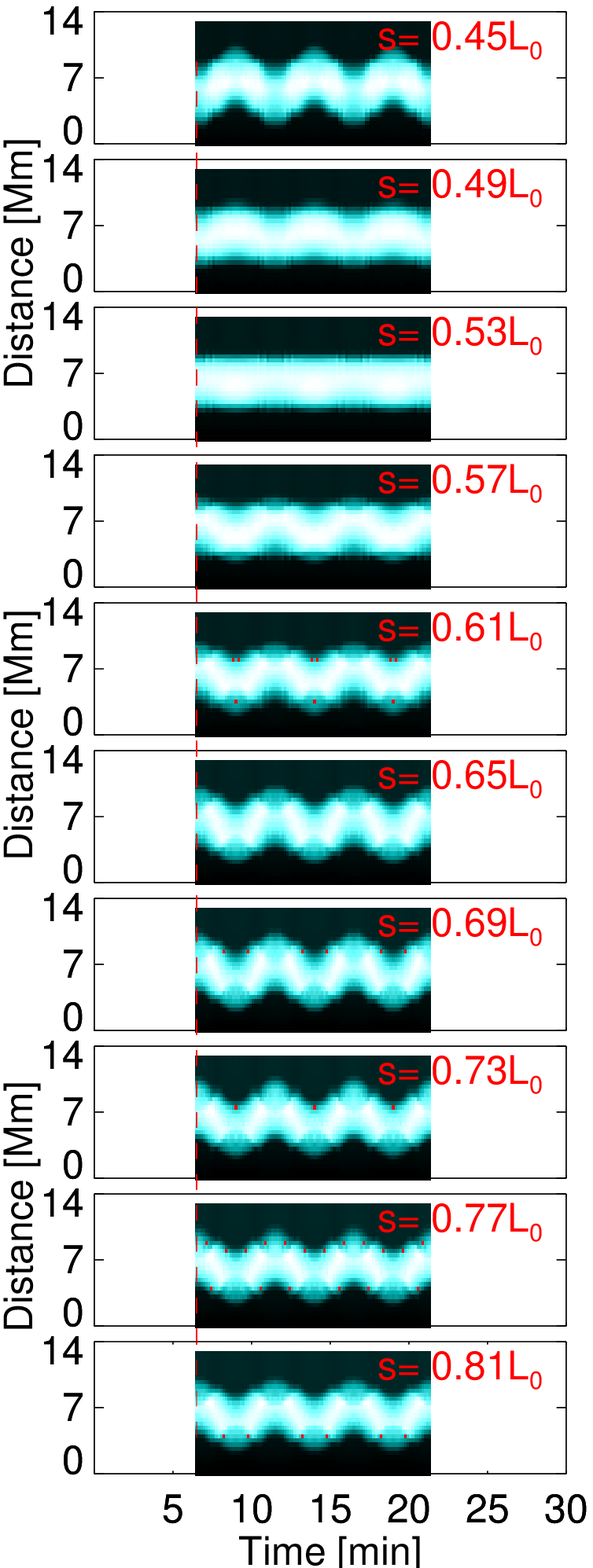}
\caption{Left panel: time distance plots along slits normal to the loop spine at
selected loop coordinates at \ew. The crosses are the traced loop motions, while
the green continuous lines are the sinusoidal fits. Middle and right panels:
same as left panel, but extracted at the synthetic loop views for the $n=3$
horizontal and vertical modes, respectively (\mw).
\label{fig:xt_w}}
\end{figure*}

\begin{figure}[ht]
\centering
\includegraphics[width=0.5\textwidth]{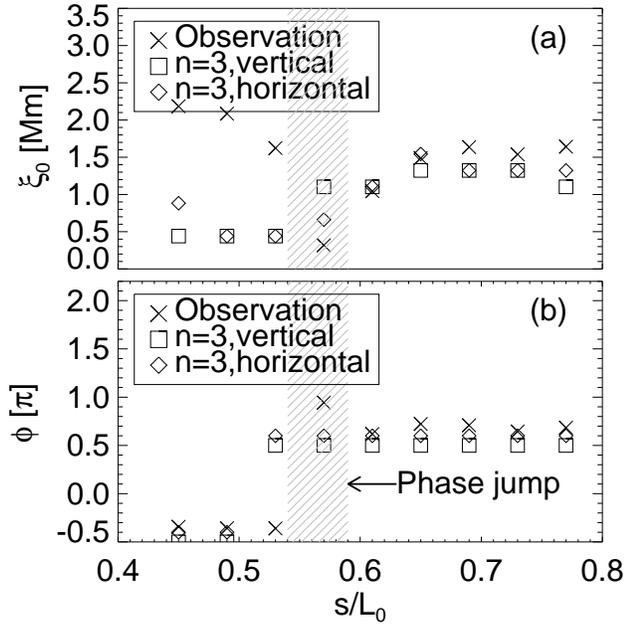}
\caption{Amplitude (a) and phase (b) of the loop oscillation as functions of
loop coordinate, measured in observation (\ew) and the $n=3$ horizontal and
vertical modes (Model $\text{W}_3$), respectively. The hatched areas highlight a
$180\deg$-phase jump. \label{fig:ph_ew}}
\end{figure}

\figref{fig:xt_w} illustrates the time distance plots extracted from the AIA 131
\AA{} observations and synthetic views of the $n=3$ horizontal and vertical
overtones. We compare the oscillation amplitude and phase distribution along the
loop coordinate and attempt to identify the loop node (\figref{fig:ph_ew}).

\begin{deluxetable}{llll}
\tablecolumns{4}
\tablewidth{0pc}
\tablecaption{List of transverse loop oscillations \label{tab:events}}
\tablehead{ Kink wave                       & \ev        &  \ea   & \ew  }
\startdata
Active region                   & AR 11283        & AR 11112    &   11121\\
 Date of observation             & 06 Sep 2011     & 16 Oct 2010 & 03 Nov 2010 
\\
 Time interval of observation    & 22:19-22:30 UT  & 19:13-19:35 UT &
12:10-12:40 UT \\
 Flare class (start time)        & X2.1 (22:12 UT) &  M2.9 (19:07 UT) & C3.4
(12:12 UT) \\
 EUV channel                                &171 \AA{} &171 \AA{}    &131 \AA{}
and 94 \AA{}\\
 Characteristic temperature $\unit{[MK]}$   & 0.6      & 0.6         & 10 \\
 Longitudinal mode number $n$                          & 1       & 1           &
  2 or 3     \\
 Polarisation: horizontal (H) or vertical (V) ?        & H       & V           &
H or V   \\
 Loop length $L_0 \unit{[Mm]}$                       & $160\pm20$& 163      & 
240     \\
 Loop radius $a\unit{[Mm]}$                          & 0.85      & $2.5\pm0.3$ &
3.8    \\
 Internal magnetic field $B_\i\unit{[G]}$            & 32-41     & $4.0\pm0.7$ &
 \nodata       \\
 Internal plasma density $\rho_\i\unit{[10^{-12}kg\, m^{-3}]}$ & 1.2   &
$0.32\pm0.05$ & 5.4    \\
 Internal electron density $n_{e\i}\unit{[10^{9}cm^{-3}]}$     & $\mtilde0.7$   
&  $0.19\pm0.03$ & 3.2       \\
 Density ratio $\rho_\i/\rho_\e$                               & 1.0-3.3       &
11-14      &   \nodata    \\
 Internal temperature $T_\i\unit{[10^6K]}$                     & 0.8        &
$0.57\pm0.14$  &  10     \\
 Internal \alfven speed $V_{\A\i}\unit{[km\,s^{-1}]}$          & 1860-2620    &
$560\pm100$ & \nodata      \\
 External \alfven speed $V_{\A\e}\unit{[km\,s^{-1}]}$          & \nodata    &
$1940\pm100$ &   \nodata    \\
 Oscillation period $P_0\unit{[s]}$                              &$122\pm6$ &
$370\pm30$ &  $302\pm14$ \\ 
 Amplitude of displacement $\xi_0\unit{[Mm]}$              & 0.9-2.9
($1.0a$-$3.4a$)\tablenotemark{*}& 1.4-2.2 (0.56a-0.88a)  &  4.7 (1.2a)  
\enddata
\tablenotetext{*}{Value in brackets indicates displacement in units of loop
radii.}
\end{deluxetable}

\begin{deluxetable}{lrrrrr}
\rotate
\tablecolumns{7}
\tablewidth{0pc}
\tablecaption{Parameters of the loop systems and the standing kink modes
\label{tab:models}}
\tablehead{Loops  & \mv    &  \ma   & Model
$\text{W}_1$ & Model
$\text{W}_2$ & Model $\text{W}_3$}
\startdata
 Loop length $L_0 \unit{[Mm]}$                     & 160        & 163          &
240    & 240  & 240  \\
 Loop radius $a\unit{[Mm]}$                        & 0.85       & 2.5          &
3.0    & 3.0  & 3.0  \\
 Internal magnetic field $B_\i\unit{[G]}$          & 30         & 4.0          &
25     & 11     & 9.0   \\
 External magnetic field $B_\e\unit{[G]}$          & 30         & 4.1          &
28    &  17     & 15 \\
 Internal plasma density $\rho_\i\unit{[10^{-12}kg\, m^{-3}]}$ & 1.67  & 0.37  &
4.2   & 4.2    & 5.0 \\
 Internal electron density $n_{e\i}\unit{[10^{9}cm^{-3}]}$     & 1.0   & 0.22  &
2.5   & 2.5    & 3.0   \\
 Density ratio $\rho_\i/\rho_\e$                    & 3.0        & 12          &
6.0  & 5.0    & 2.0  \\
 Internal temperature $T_\i\unit{[10^6K]}$          & 0.8        & 0.57        &
10   & 10  & 10  \\
 Temperature ratio $T_\i/T_\e$                      & 1.5        & 1.5         &
2.0  & 2.0    & 2.0 \\
 Internal plasma beta $\beta_i$                     & 0.0062     & 0.054       & 
0.27 & 1.4     & 2.5 \\
 External plasma beta $\beta_e$                      & 0.0014     & 0.0029     &
0.018 & 0.06    & 0.22\\
 Internal acoustic speed $C_{\s\i}\unit{[km\,s^{-1}]}$ & 150     &  130        &      
520   & 520     &520   \\
 Internal \alfven speed $V_{\A\i}\unit{[km\,s^{-1}]}$ & 2100    & 590         &
1090 & 480     &360  \\
 External acoustic speed $C_{\s\e}\unit{[km\,s^{-1}]}$ & 120     & 100        &
370      &370     &370   \\
 External \alfven speed $V_{\A\e}\unit{[km\,s^{-1}]}$ & 3600    & 2100        &
3000 &1600    &870  \\
 Longitudinal mode number $n$                         & 1       & 1           & 
1    & 2      & 3  \\
 Period $P_0\unit{[s]}$  			      &  126    & 403         &
317  & 303     & 277 \\
 Amplitude of displacement $\xi_0\unit{[Mm]}$     & 1.9 ($2.2a$)& 4.5 ($1.8a$) &
1.5 (0.5a)& 1.5 (0.5a) & 1.5 ($0.5a$) \\
 Amplitude of velocity perturbation $\hat{v}\unit{[km\,s^{-1}]}$   &100         
&70   42&   & 30  & 35   \\
 Relative amplitude of density perturbation $\hat{\rho}_1/\rho_\i$  & 0.0003    
 &0.003 & 0.0006 & 0.003 & 0.005   \\
 Relative amplitude of temperature perturbation $\hat{T}_1/T_\i$    & 0.0002    
 &0.002 & 0.0004 & 0.002 & 0.003   \\
 AIA channel                                           &171 \AA{} &171 \AA{}   
&131 \AA{} &131 \AA{}&131 \AA{}\\
 Polarisation: horizontal (H) or vertical (V) ?        & H       & H \& V       
   & H \& V  & H \& V & H \& V\\
 Viewing angle $[\tau,\eta]$                           & $[30\deg,130\deg]$ &
$[32\deg,135\deg]$  &$[90\deg, 25\deg]$&$[90\deg, 25\deg]$ &$[90\deg, 25\deg]$  \\
 Centre of the loop baseline & $[226.3\arcsec,215.3\arcsec]$ & 
$[646.2\arcsec,-274.8\arcsec]$ & \multicolumn{3}{c}{$[-884.7\arcsec,-392.9\arcsec]$} \\
 Rotation angle of synthetic image (clockwise) & $3\deg$ &$5\deg$ &$-120\deg$&$-120\deg$&$-120\deg$
\enddata

\end{deluxetable}

\section{Applications}
\label{sec:app}
\subsection{Mode identification}
\label{sec:mode}
\ev is a horizontally polarised fundamental kink mode, given the simple geometry
and projection (\figref{fig:fov_ev}). The time distance plot of the synthetic
data are consistent with the observation (\figref{fig:xt_v}): the oscillations
at different position of the loop are coherently in phase, and the emission
intensity maxima are found when the loop oscillates to the extreme positions. 

\figref{fig:fov_ma} compares the difference images of \ea and synthetic data of
the $n=1$ horizontal and vertical polarisations. The vertical mode agrees better
with the observation: it  stretches the loop geometry; and the oscillation phase
remains unchanged in this viewing angle (\figref{fig:fov_ma}{d}). In the
horizontal mode, the oscillation phase jumps by $180\deg$ at the right leg owing
to the LOS effect. The maxima of emission intensity could not be effectively
measured in observation, therefore, no further information could be extracted
from the intensity variation owing to the contamination of other loops
(\figref{fig:xt_a}).

In \ew, one leg of the loop blended into the background and could not be
effectively identified; however the rest of the loop gave a possible geometry
for the missing leg (\figref{fig:fov_rw}{a}). By comparing the difference images
of \ew and \mw, one could tell that both the $n=3$ horizontal and vertical modes
agree with the observation (\figref{fig:fov_rw}), while the $n=2$ modes do not
give the right position of the node. The left panel of \figref{fig:xt_w}
illustrates that the emission intensity reached maxima when the loop oscillated
to extreme positions; it implies that the horizontal polarisation is more likely
to be the right mode. In the follow-up analysis of this event, we, henceforth,
only consider the $n=3$ modes.  

\subsection{Amplitude and phase distribution}
\label{sec:amp}

\figref{fig:ph_ev} compares the amplitude and phase distribution of \ev and \mv.
\mv reproduces the general trend of amplitude distribution along the loop and
successfully matches the location of maximum amplitude. The phase extracted in
\mv is constant at the selected loop coordinate, while those measured in the
observation scatters around the synthetic values.  

\figref{fig:ph_ea} presents the case of \ea and \ma. The $n=1$ vertical mode
reproduces \ea better: both the general profile and the position of maximum
displacement. Again, the match between the phases of the observation and models
is excellent. However, close to the footpoint, the horizontal mode exhibits a
$180\deg$ phase jump, as also illustrated in \figref{fig:fov_ma}. Near the
footpoint, the observational errors are large in the phase, and thus, this can
not be used to distinguish the mode. 

\figref{fig:ph_ew} studies the case of \ew and \mw. Since we already determined
the longitudinal overtone (see \secref{sec:mode}), only two polarisations of the
$n=3$ modes are plotted. The amplitude finds a minimum at a node around
$s/L_0=0.54-0.59$, the horizontal mode reproduces this minimum at a close
position. The phase jump is also well modelled by both modes. By considering the
difference images (\figref{fig:fov_rw}) and the profile of the oscillation
amplitude (\figref{fig:ph_ew}{a}), we conclude that the $n=3$ horizontal mode
agrees better with the \ew than the other modes. 

\subsection{Loop intensity and width oscillations}
\label{sec:width}

In \citetalias{yuan2015sv}, we show that a quadrupole term in the kink mode could deform the loop cross-section, and thus, the loop width is liable to a periodic modulation at half of the kink mode period. In spectral observations, the non-thermal spectral line broadening, caused by the quadrupole term, leads to line intensity suppression at loop edges, so it further enhances the effective loop width modulation. In imaging observations, the line intensity suppression at loop edges does not exist, however, if the loop displacement is large enough, this effect could also be observed. In \ev, the loop of interest was clear from background contamination, and it had an displacement at the order of two loop radii. Therefore, \ev is selected to demonstrate the loop width modulation effect. We extracted the time series of the loop displacement, normalised flux and width variation at $s=0.5L_0$, and measured the oscillation period with Lamb-Scargle periodogram \citep[see, e.g.,][]{scargle1982,horne1986,yuan2011lp}. 

\begin{figure*}[ht]
\centering
\includegraphics[width=0.8\textwidth]{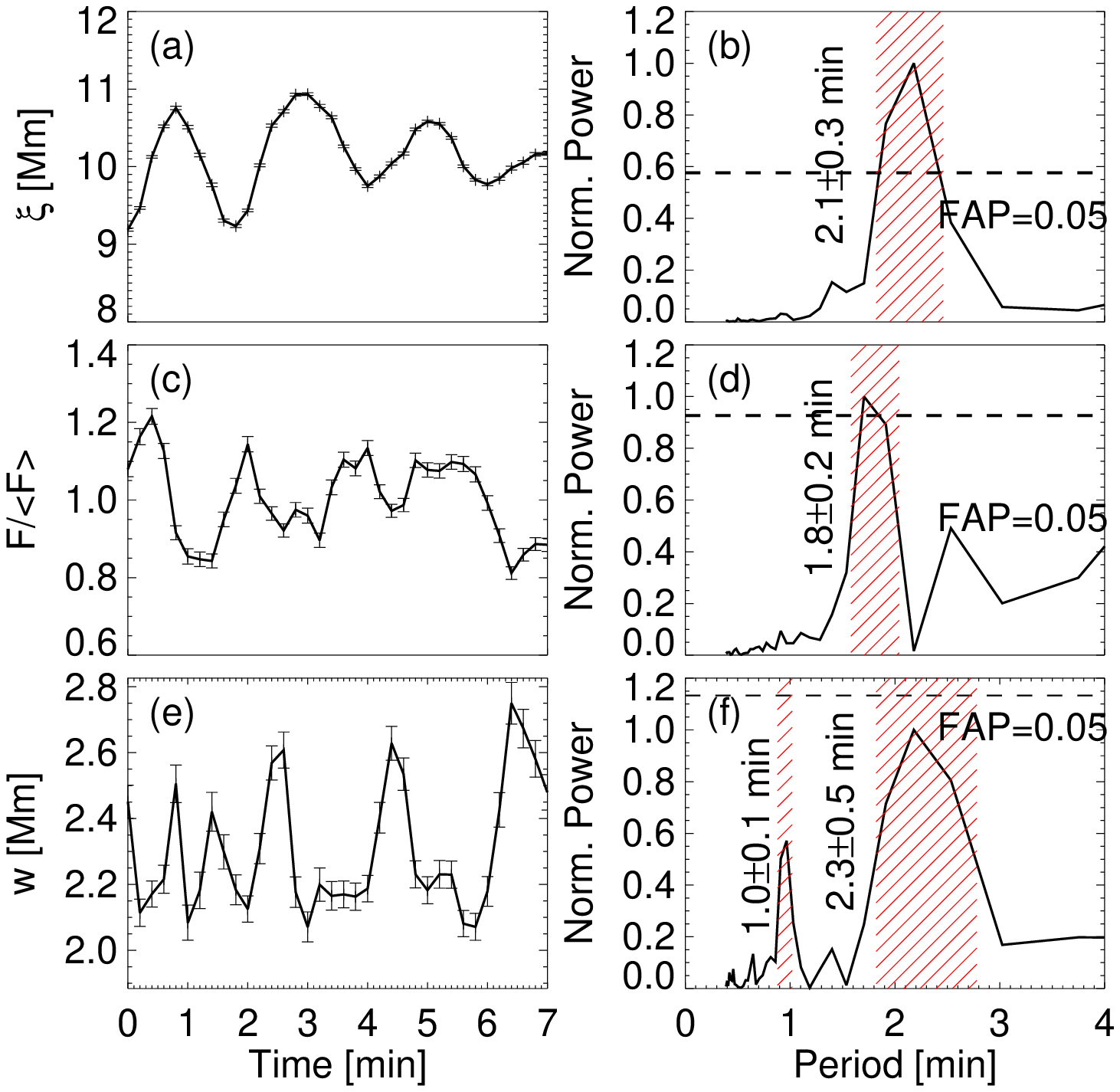}
\caption{(a),(c) and (e) are the time series of loop displacement $\xi$,
normalised emission flux $F/<F>$,and loop width $w$, measured at $s=0.5L_0$ in
\ev, respectively; while (b), (d) and (f) are the corresponding spectra. The
dashed lines mark the relevant false alarm probability (FAP) at 0.05. The
hatched areas highlight the prominent oscillation periods. \label{fig:wt_ev}}
\end{figure*}

\begin{figure*}[ht]
\centering
\includegraphics[width=0.8\textwidth]{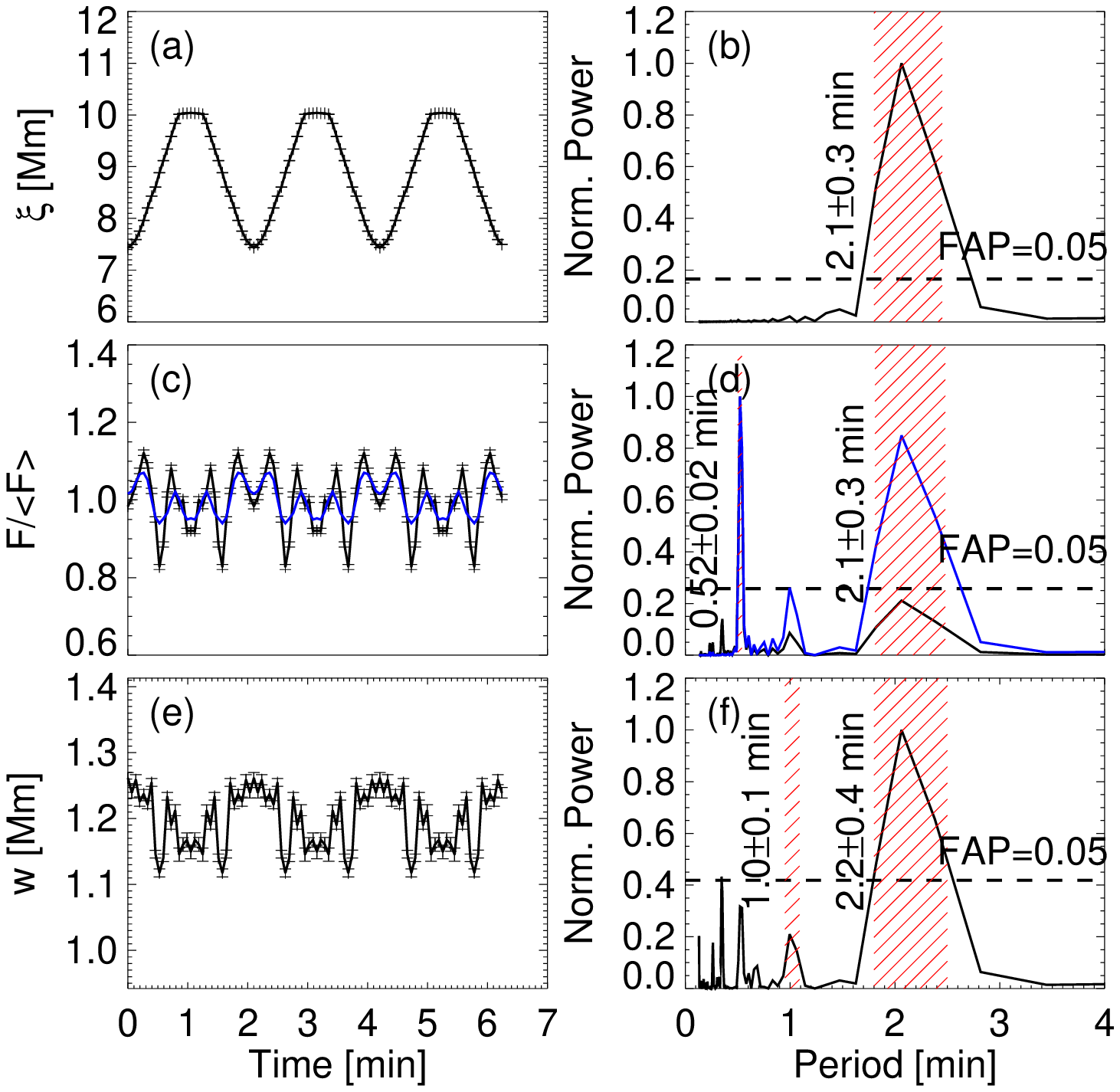}
\caption{Same as \figref{fig:wt_ev}, but for \mv. In panel (c), the blue
continuous line plots the 4-point running average of normalised flux; and the
power spectrum (blue line) has enhanced peaks in long-period range.
\label{fig:wt_mv}}
\end{figure*}

The loop oscillated back and forth about every $2\unit{min}$ with an amplitude
of about $2\unit{Mm}$ (\figref{fig:wt_ev} (a) and (b)). The modelled loop
oscillation reproduces similar amplitude and periodicity (\figref{fig:wt_mv} (a) and (b)). 

The normalized flux also has periodic variations, and the power spectrum
exhibits a prominent peak at $1.8\pm0.2\unit{min}$ (\figref{fig:wt_ev} (c) and
(d)), which has a False Alarm Probability \citep[FAP, see][for
definition]{horne1986,yuan2011lp} or a significance level less than 0.05. The
peak value is consistent with the oscillation period of the kink oscillation, if
we consider the $1\sigma$ error bar ($1.6-2.0\unit{min}$ vs $1.9-2.4\unit{min}$)
and the coarse resolution of the spectra. The modulation depth is about 20\% of
the average loop intensity. In the synthetic loop oscillation, the flux also
shows the period of the kink oscillation, but also its harmonics at
$0.52\unit{min}$ and $1.0\unit{min}$. The strongest periodicity is at
$0.52\pm0.02\unit{min}$, this may be due to the complex motions of the
quasi-rigid kink oscillations, the quadrupole terms and the breaking of symmetry
due to the LOS effect. The lack of this period in the observation may be cause
of lower time resolution, therefore, we do a four-point moving average on the
times series and re-calculate the power spectrum (blue lines in
\figref{fig:wt_mv} (c) and (d)). Now the periodicity at $2.1\pm0.3\unit{min}$
become more prominent and is more consistent with observation.  

The loop width was also measured and appears to vary with a clear periodicity.
The amplitude is about $0.15\unit{Mm}$ ($0.17a$), about 15\% of the loop
displacement. The order of magnitude is consistent with the measurement in
\citetalias{yuan2015sv}. Two peaks in the spectrum are measured at
$2.3\pm0.5\unit{min}$ and $1.0\pm0.1\unit{min}$, respectively, although they are
below the value of $95\%$-confidence level, but the periodicities are clearly
seem in the time series, albeit for only 2-3 cycles. In the synthetic data,
these two peaks are significantly measured. Other higher harmonics are also
seen. As we have much better time resolution, we are able to measure more
details of kink oscillations, which is beyond the detectability of modern
instrument. 

\section{Conclusions}
\label{sec:con}

In this paper, we demonstrated how forward models can be used to understand
observational data of transverse loop oscillations. Three events were selected
and forward modelled to reproduce the \sdo/AIA imaging observations. We have
performed mode identification to determine the oscillation polarisation and
overtone. Moreover, we measured the amplitude and phase distribution along the
loop, and the loop intensity and width oscillations, and  compared them with
observations. 

Longitudinal overtone could be identified by comparing difference images of the
observational and synthetic data, and further clues could be obtained by
identifying and matching the nodes in amplitude and phase distributions along
the loop. The polarisation could not be effectively fixed by difference images
alone. However, a key point is where the loop intensity variation reaches its
maxima. The horizontal mode finds its maxima when the loop oscillates to extreme
positions, while for the vertical mode, maxima are reached when the loop sweeps
across the equilibrium position.   

The longitudinal amplitude distribution could only be reproduced quantitatively
as a general trend. On the other hand, our models could reproduce the
longitudinal phase distribution very well for both the fundamental mode and
higher overtones. 

In our forward modelling, the loop intensity flux is found to oscillate with
multiple periodic components, which are basically the kink oscillation period
and its 2nd and 4th overtones. If the time resolution allows, the 4th overtone
could have the strongest power. However, with AIA, one may only observe the
fundamental mode and its 2nd overtone. But, if the kink oscillation period is
longer, then the 4th overtone may be resolved as well. 

For loops without background contamination, the loop width is measured to vary
periodically, at both the fundamental kink period and its 2nd overtone. Our
models also reproduce these periodicities. However, other higher overtones are
also possible to detect, if the instrument capability allows. 

Our model has reproduced many interesting features of kink oscillations, many of
them still await for rigid detection with modern instruments. Forward modelling
could assist in measuring overtone mode number, identifying polarisation,
investigating the amplitude and phase distribution, and predict the possible
origin of intensity and width variations.

\acknowledgements
The research was supported by an Odysseus grant of the FWO Vlaanderen, the IAP
P7/08 CHARM (Belspo), the Topping-Up grant CorSeis and the GOA-2015-014
(KU~Leuven), and the Open Research Program KLSA201504 of the Key Laboratory of
Solar Activity, National Astronomical Observatories of China (DY). CHIANTI is a
collaborative project involving George Mason University, the University of
Michigan (USA) and the University of Cambridge (UK).

{\it Facility:} \facility{\sdo (AIA)}

\bibliographystyle{apj}
\bibliography{yuan2015as}

\begin{thebibliography}{}
\expandafter\ifx\csname natexlab\endcsname\relax\def\natexlab#1{#1}\fi

\bibitem[{{Anfinogentov} {et~al.}(2013){Anfinogentov}, {Nistic{\`o}}, \&
  {Nakariakov}}]{anfinogentov2013}
{Anfinogentov}, S., {Nistic{\`o}}, G., \& {Nakariakov}, V.~M. 2013, \aap, 560,
  A107

\bibitem[{{Anfinogentov} {et~al.}(2015){Anfinogentov}, {Nakariakov}, \&
  {Nistic{\`o}}}]{anfinogentov2015}
{Anfinogentov}, S.~A., {Nakariakov}, V.~M., \& {Nistic{\`o}}, G. 2015, \aap,
  583, A136

\bibitem[{{Antolin} \& {Van Doorsselaere}(2013)}]{antolin2013}
{Antolin}, P., \& {Van Doorsselaere}, T. 2013, \aap, 555, A74

\bibitem[{{Arregui} {et~al.}(2007){Arregui}, {Andries}, {Van Doorsselaere},
  {Goossens}, \& {Poedts}}]{arregui2007}
{Arregui}, I., {Andries}, J., {Van Doorsselaere}, T., {Goossens}, M., \&
  {Poedts}, S. 2007, \aap, 463, 333

\bibitem[{{Aschwanden} {et~al.}(2013){Aschwanden}, {Boerner}, {Schrijver}, \&
  {Malanushenko}}]{aschwanden2013}
{Aschwanden}, M.~J., {Boerner}, P., {Schrijver}, C.~J., \& {Malanushenko}, A.
  2013, \solphys, 283, 5

\bibitem[{{Aschwanden} {et~al.}(2002){Aschwanden}, {de Pontieu}, {Schrijver},
  \& {Title}}]{aschwanden2002}
{Aschwanden}, M.~J., {de Pontieu}, B., {Schrijver}, C.~J., \& {Title}, A.~M.
  2002, \solphys, 206, 99

\bibitem[{{Aschwanden} {et~al.}(1999){Aschwanden}, {Fletcher}, {Schrijver}, \&
  {Alexander}}]{aschwanden1999}
{Aschwanden}, M.~J., {Fletcher}, L., {Schrijver}, C.~J., \& {Alexander}, D.
  1999, \apj, 520, 880

\bibitem[{{Aschwanden} {et~al.}(2003){Aschwanden}, {Nightingale}, {Andries},
  {Goossens}, \& {Van Doorsselaere}}]{aschwanden2003}
{Aschwanden}, M.~J., {Nightingale}, R.~W., {Andries}, J., {Goossens}, M., \&
  {Van Doorsselaere}, T. 2003, \apj, 598, 1375

\bibitem[{{Aschwanden} \& {Schrijver}(2011)}]{aschwanden2011}
{Aschwanden}, M.~J., \& {Schrijver}, C.~J. 2011, \apj, 736, 102

\bibitem[{{Boerner} {et~al.}(2012){Boerner}, {Edwards}, {Lemen}, {Rausch},
  {Schrijver}, {Shine}, {Shing}, {Stern}, {Tarbell}, {Title}, {Wolfson},
  {Soufli}, {Spiller}, {Gullikson}, {McKenzie}, {Windt}, {Golub}, {Podgorski},
  {Testa}, \& {Weber}}]{boerner2012}
{Boerner}, P., {Edwards}, C., {Lemen}, J., {et~al.} 2012, \solphys, 275, 41

\bibitem[{{Chen} \& {Peter}(2015)}]{chen2015}
{Chen}, F., \& {Peter}, H. 2015, \aap, 581, A137

\bibitem[{{Chen} {et~al.}(2011){Chen}, {Feng}, {Li}, {Song}, {Xia}, {Kong}, \&
  {Li}}]{chen2011}
{Chen}, Y., {Feng}, S.~W., {Li}, B., {et~al.} 2011, \apj, 728, 147

\bibitem[{{Chen} {et~al.}(2010){Chen}, {Song}, {Li}, {Xia}, {Wu}, {Fu}, \&
  {Li}}]{chen2010}
{Chen}, Y., {Song}, H.~Q., {Li}, B., {et~al.} 2010, \apj, 714, 644

\bibitem[{{De Moortel} \& {Nakariakov}(2012)}]{demoortel2012}
{De Moortel}, I., \& {Nakariakov}, V.~M. 2012, Royal Society of London
  Philosophical Transactions Series A, 370, 3193

\bibitem[{{De Moortel} \& {Pascoe}(2009)}]{demoortel2009}
{De Moortel}, I., \& {Pascoe}, D.~J. 2009, \apjl, 699, L72

\bibitem[{{Edwin} \& {Roberts}(1983)}]{edwin1983}
{Edwin}, P.~M., \& {Roberts}, B. 1983, \solphys, 88, 179

\bibitem[{{Goossens} {et~al.}(2014){Goossens}, {Soler}, {Terradas}, {Van
  Doorsselaere}, \& {Verth}}]{goossens2014}
{Goossens}, M., {Soler}, R., {Terradas}, J., {Van Doorsselaere}, T., \&
  {Verth}, G. 2014, \apj, 788, 9

\bibitem[{{He} {et~al.}(2009){He}, {Marsch}, {Tu}, \& {Tian}}]{he2009}
{He}, J., {Marsch}, E., {Tu}, C., \& {Tian}, H. 2009, \apjl, 705, L217

\bibitem[{{Horne} \& {Baliunas}(1986)}]{horne1986}
{Horne}, J.~H., \& {Baliunas}, S.~L. 1986, \apj, 302, 757

\bibitem[{{Jess} {et~al.}(2015){Jess}, {Morton}, {Verth}, {Fedun}, {Grant}, \&
  {Giagkiozis}}]{jess2015}
{Jess}, D.~B., {Morton}, R.~J., {Verth}, G., {et~al.} 2015, \ssr, 190, 103

\bibitem[{{Kumar} {et~al.}(2013){Kumar}, {Cho}, {Chen}, {Bong}, \&
  {Park}}]{kumar2013b}
{Kumar}, P., {Cho}, K.-S., {Chen}, P.~F., {Bong}, S.-C., \& {Park}, S.-H. 2013,
  \solphys, 282, 523

\bibitem[{{Lemen} {et~al.}(2012){Lemen}, {Title}, {Akin}, {Boerner}, {Chou},
  {Drake}, {Duncan}, {Edwards}, {Friedlaender}, {Heyman}, {Hurlburt}, {Katz},
  {Kushner}, {Levay}, {Lindgren}, {Mathur}, {McFeaters}, {Mitchell}, {Rehse},
  {Schrijver}, {Springer}, {Stern}, {Tarbell}, {Wuelser}, {Wolfson}, {Yanari},
  {Bookbinder}, {Cheimets}, {Caldwell}, {Deluca}, {Gates}, {Golub}, {Park},
  {Podgorski}, {Bush}, {Scherrer}, {Gummin}, {Smith}, {Auker}, {Jerram},
  {Pool}, {Soufli}, {Windt}, {Beardsley}, {Clapp}, {Lang}, \&
  {Waltham}}]{lemen2012}
{Lemen}, J.~R., {Title}, A.~M., {Akin}, D.~J., {et~al.} 2012, \solphys, 275, 17

\bibitem[{{Liu} \& {Ofman}(2014)}]{liu2014}
{Liu}, W., \& {Ofman}, L. 2014, \solphys, 289, 3233

\bibitem[{{Nakariakov} {et~al.}(2009){Nakariakov}, {Aschwanden}, \& {van
  Doorsselaere}}]{nakariakov2009}
{Nakariakov}, V.~M., {Aschwanden}, M.~J., \& {van Doorsselaere}, T. 2009, \aap,
  502, 661

\bibitem[{{Nakariakov} \& {Ofman}(2001)}]{nakariakov2001}
{Nakariakov}, V.~M., \& {Ofman}, L. 2001, \aap, 372, L53

\bibitem[{{Nakariakov} {et~al.}(1999){Nakariakov}, {Ofman}, {Deluca},
  {Roberts}, \& {Davila}}]{nakariakov1999}
{Nakariakov}, V.~M., {Ofman}, L., {Deluca}, E.~E., {Roberts}, B., \& {Davila},
  J.~M. 1999, Science, 285, 862

\bibitem[{{Nakariakov} \& {Verwichte}(2005)}]{nakariakov2005}
{Nakariakov}, V.~M., \& {Verwichte}, E. 2005, Living Reviews in Solar Physics,
  2, 3

\bibitem[{{Nistic{\`o}} {et~al.}(2013){Nistic{\`o}}, {Nakariakov}, \&
  {Verwichte}}]{nistico2013}
{Nistic{\`o}}, G., {Nakariakov}, V.~M., \& {Verwichte}, E. 2013, \aap, 552, A57

\bibitem[{{Pascoe} \& {De Moortel}(2014)}]{pascoe2014}
{Pascoe}, D.~J., \& {De Moortel}, I. 2014, \apj, 784, 101

\bibitem[{{Scargle}(1982)}]{scargle1982}
{Scargle}, J.~D. 1982, \apj, 263, 835

\bibitem[{{Schrijver} {et~al.}(2002){Schrijver}, {Aschwanden}, \&
  {Title}}]{schrijver2002}
{Schrijver}, C.~J., {Aschwanden}, M.~J., \& {Title}, A.~M. 2002, \solphys, 206,
  69

\bibitem[{{Van Doorsselaere} {et~al.}(2011{\natexlab{a}}){Van Doorsselaere},
  {De Groof}, {Zender}, {Berghmans}, \& {Goossens}}]{vandoorsselaere2011b}
{Van Doorsselaere}, T., {De Groof}, A., {Zender}, J., {Berghmans}, D., \&
  {Goossens}, M. 2011{\natexlab{a}}, \apj, 740, 90

\bibitem[{{Van Doorsselaere} {et~al.}(2011{\natexlab{b}}){Van Doorsselaere},
  {Wardle}, {Del Zanna}, {Jansari}, {Verwichte}, \&
  {Nakariakov}}]{vandoorsselaere2011}
{Van Doorsselaere}, T., {Wardle}, N., {Del Zanna}, G., {et~al.}
  2011{\natexlab{b}}, \apjl, 727, L32

\bibitem[{{Verwichte} {et~al.}(2013){Verwichte}, {Van Doorsselaere}, {Foullon},
  \& {White}}]{verwichte2013}
{Verwichte}, E., {Van Doorsselaere}, T., {Foullon}, C., \& {White}, R.~S. 2013,
  \apj, 767, 16

\bibitem[{{Wang} {et~al.}(2008){Wang}, {Solanki}, \& {Selwa}}]{wang2008}
{Wang}, T.~J., {Solanki}, S.~K., \& {Selwa}, M. 2008, \aap, 489, 1307

\bibitem[{{White} {et~al.}(2012){White}, {Verwichte}, \&
  {Foullon}}]{white2012b}
{White}, R.~S., {Verwichte}, E., \& {Foullon}, C. 2012, \aap, 545, A129

\bibitem[{{Yuan} {et~al.}(2011){Yuan}, {Nakariakov}, {Chorley}, \&
  {Foullon}}]{yuan2011lp}
{Yuan}, D., {Nakariakov}, V.~M., {Chorley}, N., \& {Foullon}, C. 2011, \aap,
  533, A116

\bibitem[{{Yuan} {et~al.}(2014{\natexlab{a}}){Yuan}, {Nakariakov}, {Huang},
  {Li}, {Su}, {Yan}, \& {Tan}}]{yuan2014lb}
{Yuan}, D., {Nakariakov}, V.~M., {Huang}, Z., {et~al.} 2014{\natexlab{a}},
  \apj, 792, 41

\bibitem[{{Yuan} {et~al.}(2015{\natexlab{a}}){Yuan}, {Pascoe}, {Nakariakov},
  {Li}, \& {Keppens}}]{yuan2015rs}
{Yuan}, D., {Pascoe}, D.~J., {Nakariakov}, V.~M., {Li}, B., \& {Keppens}, R.
  2015{\natexlab{a}}, \apj, 799, 221

\bibitem[{{Yuan} {et~al.}(2013){Yuan}, {Shen}, {Liu}, {Nakariakov}, {Tan}, \&
  {Huang}}]{yuan2013fw}
{Yuan}, D., {Shen}, Y., {Liu}, Y., {et~al.} 2013, \aap, 554, A144

\bibitem[{{Yuan} {et~al.}(2014{\natexlab{b}}){Yuan}, {Sych}, {Reznikova}, \&
  {Nakariakov}}]{yuan2014cf}
{Yuan}, D., {Sych}, R., {Reznikova}, V.~E., \& {Nakariakov}, V.~M.
  2014{\natexlab{b}}, \aap, 561, A19

\bibitem[{{Yuan} \& {Van Doorsselaere}(2015)}]{yuan2015sv}
{Yuan}, D., \& {Van Doorsselaere}, T. 2015, submitted to \apjs

\bibitem[{{Yuan} {et~al.}(2015{\natexlab{b}}){Yuan}, {Van Doorsselaere},
  {Banerjee}, \& {Antolin}}]{yuan2015fm}
{Yuan}, D., {Van Doorsselaere}, T., {Banerjee}, D., \& {Antolin}, P.
  2015{\natexlab{b}}, \apj, 807, 98

\bibitem[{{Zimovets} \& {Nakariakov}(2015)}]{zimovets2015}
{Zimovets}, I.~V., \& {Nakariakov}, V.~M. 2015, \aap, 577, A4

\end{thebibliography}

\end{document}